\documentclass[11pt]{article}
\usepackage[T1]{fontenc}
\usepackage[latin1]{inputenc}
\usepackage{amsmath,amssymb,euscript}
\usepackage{bbm}
\usepackage{graphicx}
\usepackage{epic}
\usepackage{epsfig}
\usepackage{pstricks}
\usepackage{psfrag}
 \usepackage{curves}
\usepackage{mathrsfs}
\usepackage{rotating}
\usepackage{color}

\def\be{\begin{eqnarray}}
\def\ee{\end{eqnarray}}
\def\ben{\begin{eqnarray*}}
\def\een{\end{eqnarray*}}

\numberwithin{equation}{section}
\numberwithin{figure}{section}

\def\be{\begin{eqnarray}}
\def\ee{\end{eqnarray}}

\def\me{\medskip\noindent}
\def\bi{\bigskip\noindent}

\newcommand{\Sgn}{\mbox{Sgn}}

\def\N{\mathbb{N}}

\def\R{\mathbb{R}}
\def\E{\mathbb{E}}

\title{The effect of competition and horizontal trait inheritance on invasion, fixation, and polymorphism}

\author{Sylvain Billiard\thanks{Univ. Lille, CNRS, UMR 8198 - Evo-Eco-Paleo, F-59000 Lille, France; E-mail: \texttt{sylvain.billiard@univ-lille1.fr}},  Pierre Collet\thanks{CPHT, Ecole Polytechnique, CNRS, route de
    Saclay, 91128 Palaiseau Cedex-France; E-mail: \texttt{collet@cpht.polytechnique.fr}}, R\'egis Ferri\`ere\thanks{Eco-Evolution Math\'ematique,
    CNRS
    UMR 7625, Ecole Normale Sup\'erieure, 46 rue d'Ulm, 75230 Paris, France; University of Arizona, Department of Ecology \& Evolutionary Biology, Tucson, AZ 85721, USA; E-mail: \texttt{ferriere@biologie.ens.fr}},  Sylvie
    M\'el\'eard\thanks{CMAP, Ecole Polytechnique, CNRS, route de
    Saclay, 91128 Palaiseau Cedex-France; E-mail: \texttt{sylvie.meleard@polytechnique.edu}}, Viet Chi Tran\thanks{Univ. Lille, CNRS, UMR 8524 - Laboratoire P. Painlev\'e, UFR de Math\'ematiques, F-59000 Lille, France; E-mail:
    \texttt{chi.tran@math.univ-lille1.fr}}}

\date{\today}

\begin{document}

\maketitle

\begin{abstract}
Horizontal transfer (HT) of heritable information or `traits' (carried by genetic elements, endosymbionts, or culture) is widespread among living organisms. Yet current ecological and evolutionary theory addressing HT is limited. We present a modeling framework for the dynamics of two populations that compete for resources and exchange horizontally (transfer) an otherwise vertically inherited trait. Competition influences individual demographics, affecting population size, which feeds back on the dynamics of transfer. We capture this feedback with a stochastic individual-based model, from which we derive a deterministic approximation for large populations. The interaction between horizontal transfer and competition makes possible the stable (or bi-stable) polymorphic maintenance of deleterious traits (including costly plasmids). When transfer rates are of a general density-dependent form, transfer stochasticity contributes strongly to population fluctuations. For an initially rare trait, we describe the probabilistic dynamics of invasion and fixation. Acceleration of fixation by HT is faster when competition is weak in the resident population. Thus, HT can have a major impact on the distribution of mutational effects that are fixed, and our model provides a basis for a general theory of the influence of HT on eco-evolutionary dynamics and adaptation.
\end{abstract}




\bigskip

\section{Introduction}

Horizontal transfer (HT) of biological information, such as genetic mobile elements, plasmids, or endosymbionts, is an important process in the evolution of species and the adaptation of populations across the whole tree of life \cite{ochmanetal2000,tettelinetal2008, keelingandpalmer2008, moranjarvik2010}. For example, plasmids are known to carry genetic factors that can affect their host's fitness dramatically (\emph{e.g.} \cite{stewartlevin1977,romanchuketal2014, getinoetal2015}). Horizontal transfer of plasmids plays a major role in the evolution of bacterial virulence and resistance to antibiotics,
heavy metals, and other pollutants. In eukaryotes, endosymbionts that have major physiological and ecological effects on their hosts can be transferred horizontally between individuals (e.g. Buchnera in aphids \cite{henryetal2013, gonellaetal2012}). Plasmids and endosymbionts can bring fitness benefits to their hosts \cite{oliveretal2008}; they also come with fitness costs, such as reduced reproduction rate or increased predation risk \cite{baltrus2013,jaenike2012}. Given their potential costs and benefits, how mobile elements such as plasmids and endosymbionts evolve and persist is not yet fully understood, and we only have limited theory to predict the conditions under which stable polymorphic populations of hosts carrying or not the mobile elements should be expected (e.g. the giant plasmids in \emph{Pseudomonas syringae} \cite{romanchuketal2014}  or the inherited facultative symbiont of the pea aphid \emph{Hamiltonella defensa} \cite{oliveretal2008}, see other examples in \cite{slateretal2008}).
\par The impact of HT on ecological dynamics has been studied with two different types of models. A first class of models was spawned in the late 1970's  by the seminal
contribution of Anderson and May on host-pathogen population dynamics \cite{andersonmay1979}. The primary motivation for these models was to investigate whether and how a pathogen might drive its host to extinction. These models highlighted the importance of whether horizontal transmission depends upon the density versus frequency of infected hosts.

\par Models have been constructed in the May-Anderson's framework specifically to investigate the impact of  HT on the dynamics of plasmids in bacteria (e.g. \cite{stewartlevin1977, levinetal1979, lili}). These models face several limitations: i) All models derived in the Anderson-May framework, and especially those concerned with plasmid transfer, are deterministic; however, even in an ideal constant environment, demographic stochasticity can be a strong influence of population persistence, invasion, and coexistence, and thus a critical factor of evolution. ii) The models make highly simplified ecological assumptions, especially regarding competition --competition between hosts is at best highly simplified, or ignored altogether. iii) In models addressing plasmid transmission, only density-dependent transmission rates have been considered, even though we know from the general host-pathogen framework that the transmission regime (density- vs frequency-dependence) can affect the population dynamics dramatically  \cite{andersonmay1979, bootssasaki2003}. In general, what type of transfer best describes a given system is not trivial, but it seems clear that
density-dependent transfer rates do not generally fit the case of plasmids \cite{stewartlevin1977, slateretal2008, getinoetal2015}.

\par A second class of models was developed in a population genetics framework to address the effect of HT on the fixation of beneficial mutations
\cite{cavalli-sforzafeldman, novozhilovetal2005, tazzymanbonhoffer}) and on the evolution of specific traits (e.g. cooperation \cite{dimitriuetal2014,
mcgintyetal2013}). \cite{baumdickerpfaffelhuber} also took a population genetics approach to HT. All these models make strong simplifying assumptions on the  ecology (e.g., competition) that further restrict their representation of transfer: there is no explicit demography and population size is kept constant thus obliterating the important distinction between frequency and density dependent transfer.

\par Finally to our knowledge, all models developed so far assume HT to be unilateral; in these models, the population is subdivided into two classes, the class of donors (e.g. infected hosts or plasmids carriers) and the class of recipients (e.g. susceptible hosts, bacteria devoid of plasmids), with HT occurring unilaterally from the former to the latter. While this assumption is certainly justified in the case of horizontal transfer of symbionts or plasmids, it is not  general. If we consider that the population is subdivided in two classes of individuals with different traits, then one can expect in general that HT occurs in both directions between the two classes, and not necessary symmetrically.

\par Our aim is to relax most of the previous limitations, by developing a mathematically rigorous  stochastic model of population dynamics with both vertical transmission and HT of traits. We consider a population of individuals undergoing a basic process of birth, interaction, and death. Interactions among individuals  are ecological (competition) and also drive the transfer of information which otherwise is inherited vertically. Our modeling framework is individual-based and stochastic; population size and dynamics are thus emergent properties that the model predicts rather than assumes. We call 'trait' an entity such as an allele, a mobile genetic element, or a plasmid, which can be inherited vertically and also  transferred horizontally upon contact between individuals.\\

\par We focus on the simplest case of two traits, A and a, and address three general questions: In large populations, what are the deterministic dynamics of the subpopulations of traits $A$ and $a$? How does the stochasticity of HT, relative to the demographic stochasticity of the birth-death process, contribute to the population fluctuations around its deterministic equilibrium? When one trait is initially rare in the population (e.g. a mutation of the common trait), how does HT influence its probability of invasion and time to fixation? Our theory covers both cases of frequency- and density-dependent HT rates; these dependencies appear as special cases of  a more general form of transfer rate, that we call Beddington-DeAngelis by analogy with a similar model used to describe contact between predators and prey (\cite{beddington1975, deangelisetal1975}).

\section{ A general stochastic model of two-type population dynamics with horizontal transfer}\label{section:modele}

We consider a population of clonal haploid individuals. The population changes in continuous time with the random occurrences of birth and death. We assume that the initial population size is of order $K$. $K$ will be used to scale individual-level parameters when we examine the case of large system size ($K\rightarrow \infty$).
We denote $N^{u,K}_t$ the number
of individuals with trait $u\in \{A,a\}$ at time $t$. The population size is also scaled by parameter $K$, and we define population \emph{densities} of traits A and a, respectively, as
$$(X^K_{t}, Y^K_{t}) = \frac{1}{K}(N^{A,K}_t, N^{a,K}_t).$$

\paragraph{Modeling death, competition and reproduction processes}

An individual with trait $u\in \{A,a\}$ gives birth to an individual with trait $u$ at rate $b_K(u)$. The death rate of an individual with trait $u$ at time $t$ is
$$d_K(u) + C_{K}(u,A) N^{A,K}_{t} +C_{K}(u,a) N^{a,K}_{t}=d_K(u) + K C_{K}(u,A)X^K_{t} + K C_{K}(u,a) Y^K_{t},$$
where $C_K(u,v)$ measures the effect of competition from an individual with trait $v$ on an individual with trait $u$. $d_K(u)$ is the natural death rate; competition adds a logistic death term, proportional to the density $X^K_t$ or $Y^K_t$ of competitors. We will denote $r_K(u)=b_K(u)-d_K(u)$
the natural growth rate of the sub-population of trait $u$ in the absence of competition.

\paragraph{Horizontal transfer}

The trait ($A$ or $a$) can be horizontally transferred between individuals. We assume that when a transfer occurs, the recipient $v$ acquires the trait $u$ of the donor $(u\neq v)$. This occurs for instance during bacterial conjugation where the donor transmits a copy of her plasmid to individuals devoid of plasmid.
Transfers can occur in both directions: from individuals $A$ to $a$ or the reverse, possibly at different rates. In a population with renormalized quantities of traits $A$ and $a$ given by $(X^K_{t}, Y^K_{t})=(x,y)\in \big({\mathbb{N}\over K}\big)^2$, a donor transfers its trait $u$ to a recipient with trait $v$  at
rate
$h_{K}(u,v, x, y)$. In the special case of bacterial conjugation and plasmid transfer (see Section 4), HT is unilateral only and
 some empirical data  suggest that the HT rate is not constant and depends on population densities \cite{stewartlevin1977}.

\paragraph{Generator of the stochastic process}

The process $(X^K_t, Y^K_t)_{t\geq 0}$ is fully described by its infinitesimal generator applied to continuous bounded test functions $F$ from $\R^2$ to $\R$. For $(x,y) \in (\N / K)^2$,
\begin{multline}
LF(x,y)=  K\, b_K(A)  x\Big[F\big(x+\frac{1}{K},y)-F(x,y)\Big]+  K\,  b_K(a)y\Big[F\big(x,y+\frac{1}{K}\big)-F(x,y)\Big]\label{generator}\\
\begin{aligned} 
 & + K\,\big(d_{K}(A)+K C_{K}(A,A)\, x+ K C_{K}(A,a)\, y\big) \, x\ \Big[F\big(x-\frac{1}{K},y\big)-F(x,y)\Big]\\
 &  + K\,\big(d_{K}(a)+ K C_{K}(a,A)\, x+ K C_{K}(a,a)\, y\big)\,y\ \Big[F\big(x,y-\frac{1}{K}\big)-F(x,y)\Big]\\
 &  + K^2\,h_K(A,a, x, y)\,x\,y\, \Big[F\big(x+\frac{1}{K}, y-\frac{1}{K}\big)-F(x,y)\Big] \\
 &  + K^2\,h_K(a,A, x, y)\,x\,y \,\Big[F\big(x-\frac{1}{K}, y+\frac{1}{K}\big)-F(x,y)\Big].
\end{aligned}
\end{multline}
The first two terms of \eqref{generator} refer to the birth of an individual with  trait $A$ or $a$. The next two  terms  refer to the death of an individual with trait $A$ or $a$,  due to natural death or competition for
resources. The last two terms refer to the change of the trait of an individual $a \rightarrow A$ or $A \rightarrow a$ due to
HT. Note that the competition and transfer terms are of order $K^2$, which is the order of the number of interacting pairs of individuals in a population of size $K$. \\

When $K \rightarrow \infty$, the terms in square brackets of \eqref{generator} are all to be approximated at order $1/K$. As a consequence we need all factors in front to be kept at order $K$. For the birth and death terms, this is the case  provided that $b_{K}(.) \to b(.)$, $d_{K}(.)\to d(.)$ and $K C_{K}(.,.) \to C(.,.)$. For the HT terms, we need  $K h_K$ to go to a finite limit. This limit depends on alternate assumptions  about the mechanism of transfer. We obtain different limits for a transfer rate that is either density-dependent, or frequency-dependent, or a mixture of both.

\paragraph{How to model different modes of horizontal transfer at the individual level}

We consider the following general form, for donor $u$ and recipient $v$ in a population $(x, y) \in \big({\mathbb{N}\over K}\big)^2$, obtained as an interpolation between frequency (\eqref{eq:freqdeprate} below) and density-dependent rates (\eqref{eq:densdeprate} below):
\begin{equation}
h_{K}(u,v, x, y)= \frac{ \psi_K(u,v) }{1+E_K(u,v) \psi_K(u,v)\  (x+y)}.\label{eq:generalrate}
\end{equation}
The latter expression highlights two components of the transfer process: pairs form at rate $ \psi_K(u,v)$ and the mean time of transfer, once a pair is formed, is $E_K(u,v)$ (see \cite{geritzgyllenberg} for a related treatment of functional response in a deterministic framework). \\

\noindent \emph {Case 1: Frequency-dependent transfer rate (FD).}  We recover a frequency-dependent transfer rate by assuming  $\lim_{K\rightarrow +\infty}\frac{E_K(u,v)}{K}=E(u,v)$, and $\lim_{K\rightarrow +\infty} K \psi_K(u,v)= + \infty$, which yields
\begin{equation}
\lim_{K\rightarrow +\infty}K h_{K}(u,v, x, y)= \frac{1}{E(u,v)(x+y)}.\label{eq:freqdeprate}
\end{equation}
Thus, frequency-dependent transfer can arise  when pair formation is very fast compared with Cases 2 and 3 (the order of $\psi_K(u,v)$ is greater than $1/K$), and when transfer events take times of order $K$.  Longer transfer times in larger population may result from
  interference between pairs and  the large number of surrounding individuals.

\me \emph {Case 2: Density-dependent transfer rate (DD).}  Assuming $\lim_{K\rightarrow +\infty}\frac{E_K(u,v)}{K }= 0$ and $\lim_{K
\rightarrow +\infty}K \psi_K(u,v)= \psi(u,v) \neq 0$, we recover the density-dependent transfer rate
\begin{equation}
\lim_{K\rightarrow +\infty}K h_{K}(u,v, x, y)= \psi(u,v).\label{eq:densdeprate}
\end{equation}
Thus, density-dependent transfer can arise when the pair formation rate is of order $1/K$ and transfer events take  time in $o(K)$, thus a negligible time. The number of pairs formed per unit time is thus proportional to $x y$.

\me \emph{Case 3: Beddington-DeAngelis like transfer rate (BDA).} Assuming  $\lim_{K\rightarrow +\infty}\frac{E_K(u,v)}{K}= E(u,v)$, and $\lim_{K
\rightarrow +\infty}
K \psi_K(u,v)=\psi(u,v) \neq 0$, we obtain the so-called Beddington-DeAngelis response form for the transfer rate:
\begin{equation}
\lim_{K\rightarrow +\infty}K h_{K}(u,v, x, y)= \frac{\psi(u,v) }{1+ E(u,v) \psi(u,v)\,  (x+y)}.\label{eq:MMrate}
\end{equation}

\section{Large population limit and the impact of horizontal transfer on the maintenance of polymorphism}\label{section:ODE}

We look at deterministic approximations of the stochastic population dynamics on the `ecological' timescale of births, interactions (competition and transfer), and deaths. This allows us to investigate how HT might affect the coexistence of traits $A$ and $a$, and the conditions of invasion of a trait that is rare in a population with the other trait. With this aim in view, we consider the following scalings: the initial population sizes are such that $(X^{K}_0,Y^K_0)\rightarrow (x_0,y_0)\in \R^2_+$ in probability and $b_K(u) \rightarrow b(u)$, $d_K(u)\rightarrow d(u)$, $\,K C_{K}(u,v)  \rightarrow {C(u,v)}$ and $\,\lim_{K\to \infty} K h_{K}(A,a,x,y)=h(A,a,x,y)$. We take
\begin{equation}
h(A,a,x,y)= \frac{\tau(A,a)}{\beta + \mu\,(x+y)},\label{tauxlimite}
\end{equation}
which covers all cases FD, DD and BDA.  For $\,\beta=1, \mu=0\,$ or $\,\beta=0, \mu=1$, one gets cases DD and FD, respectively. The BDA rate is obtained with
$\psi(u,v)=\beta^{-1} \tau(u,v)$ and $E(u,v)=\mu \tau(u,v)^{-1}$. \\

Below we show that the behavior of the deterministic dynamical system is influenced by
 HT only through the `horizontal flux' rate $$\alpha(A,a)=\tau(A,a)-\tau(a,A).$$ The horizontal flux rate quantifies the asymmetry between transfers in either direction and can be positive as well as negative (or zero in the case of perfectly symmetrical transfer). In the subsequent section we will show that the fully stochastic population process depends not only on the flux $\alpha$ but also on  $\tau$ itself.

\paragraph{Deterministic approximations and stability analysis.}

When $K \rightarrow \infty$ the sequence of stochastic processes $(X^K_., Y^K_.)_{K\in \N^*}$ converges in probability to the unique
solution of the following
system $(x_., y_.)$ of ordinary differential equations (ODEs):
\begin{align}
\frac{dx}{dt}= & \Big(r(A)-C(A,A)x-C(A,a)y+ \displaystyle{\alpha(A,a)\over \beta + \mu\,(x+y)}\,y \Big)x  \nonumber\\
\frac{dy}{dt}= & \Big(r(a)-C(a,A)x-C(a,a)y- \displaystyle{\alpha(A,a)\over \beta + \mu\,(x+y)}\,x \Big)y ,\label{eq:edo}
\end{align}
where $\, r = b - d$.

\begin{figure}[!ht]
  \includegraphics[angle=0,scale=0.7]{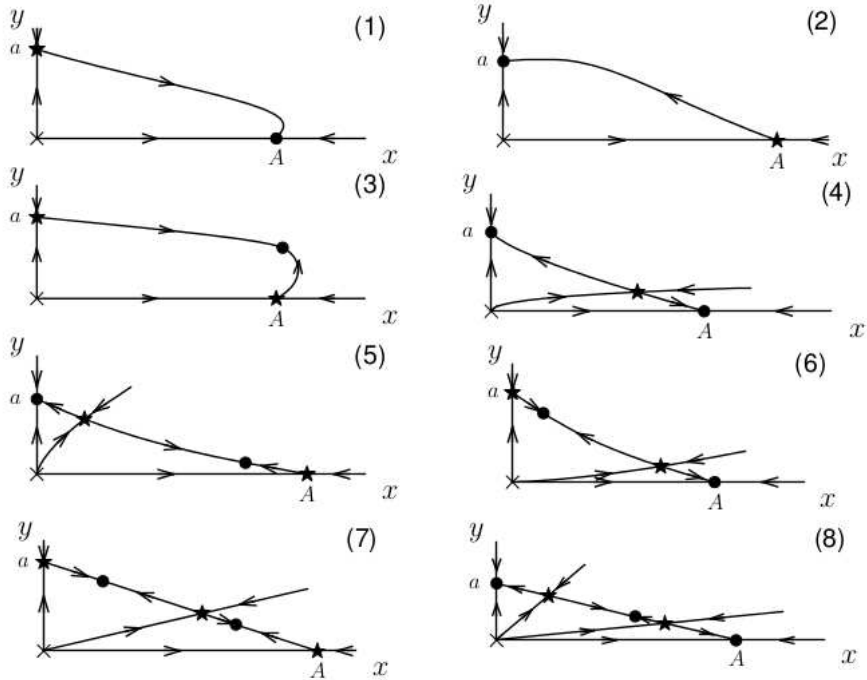}
 \caption{{\small \textit{Deterministic dynamics of a two-trait population with HT. $x$ and $y$ denote the densities of each type. Circles and stars respectively indicate stable and saddle fixed points. See text for more details.}}}\label{fig:diagram}
\end{figure}

\me Figure \ref{fig:diagram} shows  eight possible phase diagrams for the dynamical system \eqref{eq:edo}, where the circles and stars indicate stable and unstable fixed points, respectively. In theory some other phase diagrams might be possible, but have never been observed numerically. In the case where $A$ and $a$ are sufficiently similar, the only possible phase diagrams are those of Figure \ref{fig:diagram} (see mathematical proofs in the Electronic Supplementary Materials (ESM) \cite{billiardcolletferrieremeleardtran:ESM}).

The phase diagrams in Figure \ref{fig:diagram}  show that both stable polymorphic or monomorphic populations are possible, depending on the parameter values and the form of HT rates. The boundary fixed points on the $x$- and $y$-axes  correspond to the monomorphic populations of $A$ and $a$, respectively.  The dynamics close to the $y$-axis are driven by the so-called invasion fitness, denoted by $S(A,a)$, of individuals with trait $A$ in a resident population of trait $a$. A fixed point on the $x$- or $y$-axis is stable against invasion by the alternative type if the associated invasion fitness is negative; it is unstable if the invasion fitness is positive. Standard stability analysis of the boundary equilibria yields
\be S(A,a) &=&
r(A) + \left(\frac{\alpha(A,a)}{\beta+\mu \bar{y}}- C(A,a)\right)\,\overline{y} \quad  \hbox{ with } \quad \overline{y}={r(a)\over C(a,a)}\nonumber \\
&=&  f(A,a)+ {\alpha(A,a) r(a)\over
\beta C(a,a)+\mu r(a)},\label{fit}
\ee
where $f(A,a)=r(A)-\frac{C(A,a)r(a)}{C(a,a)}$ is the invasion fitness of $A$ in a resident population $a$ \emph{in the absence of} HT. Equation \eqref{fit} thus shows that HT can revert the direction of selection (i.e. $S(A,a)$ and $f(A,a)$ have opposite signs) provided 1) invasion fitness $f(A,a)$ and transfer flux rate $\alpha(A,a)$ have opposite sign, and 2) $|f(A,a)| < {r(a)\over
\beta C(a,a)+\mu r(a)} |\alpha(A,a) |$. Condition 2 is facilitated if $r(a)/(\beta C(a,a)+\mu r(a))$ is larger, which happens if the resident $a$ equilibrium population density $\overline{y}=r(a)/C(a,a)$ is large.

\paragraph{The effect of horizontal transfer on deterministic equilibria}

Compared to the classical Lotka-Volterra model, \emph{i.e.} without HT ($\alpha=0$), four new phase diagrams are possible, shown in Figures \ref{fig:diagram} (5)-(8). Figures \ref{fig:diagram} (5)-(8) show that the final state of the population strongly depends on the initial conditions: the dimorphic population state can be stable, or the population can fix one of the two traits. Interestingly, not all phase diagrams can be obtained for every form of transfer rates. In the DD case,  Figures \ref{fig:diagram} (1)-(4) are the only possibilities. In the FD case, one can also have   Figures \ref{fig:diagram} (5)-(6)  while Figures \ref{fig:diagram} (7)-(8) are not possible. For BDA, all cases  are possible (proof in ESM \cite{billiardcolletferrieremeleardtran:ESM}).

In a classical Lotka-Volterra competition model without HT, both types $A$ and $a$ can coexist if and only if their invasion fitnesses are positive. Except in the density-dependence case, our analysis of HT reveals a new picture, since stable polymorphic states can exist whatever the sign of the two invasion fitnesses. More precisely, in cases (5) and (6), one type invades the other but the latter does not invade the former, and yet both types can coexist at a stable equilibrium. In case (7) there is mutual invasibility and coexistence occurs at either one of two possible equilibria, depending on which type was common when the alternate type entered the population. In case (8) neither type is invadable, and yet stable coexistence is possible provided the initial mix contains enough of both types.

\paragraph{Special case 1: Constant  competition} Constant competition means  $C(u,v)\equiv C$ for all $u,v\in \{A,a\}$. In this case, it is easy to show that
$$f(A,a) = r(A) - r(a).$$
Expressing \eqref{eq:edo} in terms of the size of the population $n=x+y$ and proportion of trait $A$, $p=x/n$ gives:
\begin{align}
\frac{dn}{dt}= & n\,\big(p\,r(A) + (1-p)\,r(a)  - Cn\big) \nonumber\\
\frac{dp}{dt}= & p\,(1-p)\,\Big( r(A) - r(a)  + \alpha(A,a){ n\over \beta +\mu n} \Big).\label{eq:edo2-Cconstant}
\end{align}

\emph{Frequency-dependent horizontal transfer rate.} With $\,\beta=0\,$ and $\,\mu=1\,$, \eqref{eq:edo2-Cconstant}
shows that there are only two equilibria for the second equation: $\,p=0\,$ or $\,p=1\,$ (Figures \ref{fig:diagram} (1)-(2)). Therefore there is no polymorphic fixed point and we get a very simple ``Invasion-implies-Fixation'' criterion:
trait $A$ will invade a resident population of trait $a$ if and only if
\be\label{eq:FDpolym}  S(A,a)&=& f(A,a) + \alpha(A,a) = - S(a,A)>0.
\ee
Thus, compared to a system without HT, horizontal transfer can revert the direction of selection (\emph{i.e.} $S(A,a)$ and $f(A,a)$ have opposite signs) provided that
$$|\alpha(A,a)| > |f(A,a)|\quad \mbox{ and } \quad \Sgn(\alpha(A,a)) = -\Sgn(f(A,a)).$$
This underscores that HT can drive a deleterious allele to fixation, even though the population dynamics are deterministic and there is no frequency-dependent selection (for $C$ is constant).\\

\emph{ Density-dependent or BDA horizontal transfer rate.} When $\beta\not=0$, there exists a polymorphic fixed point when
\begin{equation}\label{eq:DDpolym}
0<\widehat{p}=-\frac{f(A,a)(\beta C+ \mu r(a)) +\alpha(A,a) r(a) }{\mu f(A,a)^2+ \alpha(A,a)\,f(A,a)}<1.
\end{equation}If $f(A,a)$ and $\alpha(A,a)$ are both positive, the above expression is negative and there is fixation of $A$. If $f(A,a)$ and $\alpha(A,a)$ are both negative, $\widehat{p}<1 \Leftrightarrow -f(A,a) \beta C<r(A)(\mu f(A,a)+\alpha(A,a))$ which never happens since the right hand side is positive and the left hand side is negative. So there is fixation of $a$ in this case.
When $f(A,a)$ and $\alpha(A,a)$ have opposite sign, there may exist a non-trivial fixed point
which is stable if
\be
\label{trade}\mu f(A,a)+\alpha(A,a)>0.\ee

In contrast to the classical Lotka-Volterra competion model in which constant competition prevents stable coexistence, HT with DD or BDA transfer rates allows the maintenance of a deleterious trait ($f(A,a) < 0$) in a stable polymorphic state; this requires that the flux rate ($\alpha(A,a)$) be positive and large enough in favor of $A$ to $a$. This result is reminiscent of a deleterious mutation being maintained by mutation-selection balance in a large population (i.e. with drift being negligible), with the role of `mutation' here being played by transfer - a major difference being, however, that production of new deleterious-trait individuals by transfer requires contact, the likelihood of which decreases when the density of deleterious types becomes lower.

\paragraph{Special case 2: Traits $A$ and $a$ have nearly equal phenotypic effects} We now consider the case where types $A$ and $a$ have similar phenotypic effects, such that variation between $a$ and $A$ brings only little changes in ecological parameters, assuming FD and BDA modes of HT rates (\emph{i.e.} $\mu\neq 0$). We assume that $r(a)=r$, set $C(a,a)=C$ and recall that by definition $\alpha(a,a)=0$. We assume that there is a small $\varepsilon>0$ such that
$r(A) = r + \kappa\,\varepsilon$, $C(A,a) = C + d_{1}\,\varepsilon$, $C(a,A) = C + d_{2}\,\varepsilon$, $C(A,A)= C + d_{1}\,\varepsilon + d_{2}\,\varepsilon$,
$\alpha(A,a) =\lambda\,\varepsilon.$ We also assume the parameter values $\kappa$, $d_1$, $d_2$, $\lambda$ to be drawn uniformly in $[-1,1]$. \\

Under FD horizontal transfer rates ($\beta =0$), only cases (1)-(6) can occur. We fall in cases (1)-(4) of Fig.\ref{fig:diagram} with probability of order $1- {\cal O}(\varepsilon^4)$ and with probability  $ {\cal O}(\varepsilon^4)$ in cases (5)-(6). In contrast, under BDA horizontal transfer rates ($\beta\neq 0$), all cases (1)-(8) can occur. With probability of order $1- {\cal O}(\varepsilon^4)$, we fall in cases (1)-(4) of  Fig. \ref{fig:diagram}, with probability  ${\cal O}(\varepsilon^4)$ in cases (5)-(6), with probability  ${\cal O}(\varepsilon^7)$  in case (7), and with probability smaller than ${\cal O}(\varepsilon^8)$ in case (8) (see ESM \cite{billiardcolletferrieremeleardtran:ESM}). This shows that if trait $A$ is introduced in a population by mutation with small phenotypic effects, then HT will most likely not affect the dynamics in comparison to a classical Lotka-Voltera model. Note, however, that this does \emph{not} imply that cases (5)-(8) will never occur due to their extremely small likelihood. In fact, a trait substitution sequence (whereby a sequence of mutation and selection events govern changes in the trait value, hence in the values of $\kappa$, $d_1$, $d_2$, and/or $\lambda$) may well drive the trait towards a value where (5), (6), (7), or (8) happens. Thus, a trait value around which one of the scenarios (5)-(8) occurs may be observed with a non-negligible probability if that trait value is an attractor for some set of trait substitution sequences. \\



\section{Population stochastic fluctuations due to demographic stochasticity and  stochasticity of horizontal transfer}\label{sec:fluctuation}

Even if the environment is constant, as we assume throughout this study, stochastic fluctuations in the size of the sub-populations of trait $A$ and $a$ are expected from the demographic stochasticity inherent to the individual processes of birth and death, and from the stochasticity of the HT process. Hereafter we evaluate the latter and compare the contributions of demographic stochasticity and transfer stochasticity to population fluctuations.

The Central Limit Theorem and diffusion theory allow us to study  population fluctuations on two different timescales, although the limits take similar forms: the ecological timescale of the dynamical system that we derived in the previous section; and a much longer timescale which is relevant when the population is `almost critical', \emph{i.e.} when the intrinsic growth rates and competition coefficients of both subpopulations are very close to zero, which causes the  population only to change on a very slow  time scale.

\subsection{Stochastic fluctuations around the deterministic population dynamics}

We consider the following parameter scaling, such that the convergence to the dynamical system \eqref{eq:edo} holds (see Section \ref{section:ODE}): $b_K(u)=b(u)$, $d_K(u)=d(u)$, $C_K(u,v)=C(u,v)/K$ and $K h_K(u,v,x,y)=\tau(u,v)/(\beta+\mu(x+y))$.
To gain insights into the magnitude of the stochastic fluctuations around the deterministic dynamics, we use the central limit theorem associated to the convergence of $(X^K,Y^K)_{K\in \N^*}$ to the deterministic  solution of \eqref{eq:edo}. For this we introduce the sequence $\eta^K_.=(\eta^{K,A}_.,\eta^{K,a}_.)=\big(\sqrt{K}(X^K_.-x_., \
Y^K_.-y_.)\big)_{K\in \N^*}$, where $(x_.,y_.)$ is the solution of the ODE \eqref{eq:edo}. When $K \rightarrow \infty$, $\eta^K_.$ converges in distribution, to the unique (continuous)
solution of the stochastic differential equations
\begin{align}
\eta^A_t= & \eta^A_0+ \int_0^t \Big[\Big(r(A)-2 C(A,A)x_s-C(A,a)y_s + \alpha(A,a) \frac{\beta y_s + \mu y^2_{s}}{(\beta+\mu(x_s+y_s))^2}\Big)\eta^A_s \nonumber\\
  & \hspace{0.5cm}+\Big(- C(A,a)   x_s + \alpha(A,a)  \frac{\beta x_{s} + \mu x_s^2}{\big(\beta+\mu(x_s+y_s)\big)^2}\Big)\eta^a_s \Big] ds\nonumber\\
  & \hspace{0.5cm}+ \int_0^t \sqrt{\big(b(A)+d(A)+C(A,A)x_s+C(A,a)y_s)\big)x_s}\ dW^A_s\nonumber\\
  & \hspace{0.5cm}+\int_0^t\sqrt{\frac{(\tau(A,a)+\tau(a,A))x_s y_{s}}{\beta +\mu(x_s+y_s)}}\ dW^h_s\nonumber\\
\eta^a_t= & \eta^a_0+ \int_0^t \Big[\Big(r(a)- C(a,A)x_s-2C(a,a)y_s -\alpha(A,a) \frac{\beta x_{s}+ \mu x^2_{s}}{(\beta+\mu(x_s+y_s))^2}\Big)\eta^a_s \nonumber\\
  & \hspace{0.5cm}-\Big( C(a,A)   y_s + \alpha(A,a) \frac{\beta y_s + \mu y^2_{s}}{(\beta+\mu(x_s+y_s))^2}\Big)\eta^A_s \Big] ds\nonumber\\
  & \hspace{0.5cm}+\int_0^t \sqrt{\big(b(a)+d(a)+C(a,A)x_s+C(a,a)y_s)\big)y_s }\ dW^a_s\nonumber\\
  & \hspace{0.5cm}-\int_0^t\sqrt{\frac{(\tau(A,a) +\tau(a,A))x_s y_{s}}{\beta +\mu(x_s+y_s)}}\ dW^h_s
  \label{fluct}
\end{align}
where $W^A$, $W^a$ and $W^h$ are three independent Brownian motions. These equations show that the deterministic dynamical system \eqref{eq:edo} approximates $(X^K,Y^K)$ with an error of order $1/\sqrt{K}$. The first terms $\eta^A_0$ and $\eta^a_0$ correspond to the initial conditions. The fluctuations can be decomposed into two random terms. The integral in $ds$ varies regularly with time (as a predictable finite variation process) whereas the stochastic integrals with respect to the Brownian motions are additional irregular Gaussian noises. Notice that the first stochastic integrals in $W^A$ and $W^a$ correspond to the stochastic fluctuations due to birth and death, whereas the integrals in $W^h$ correspond to fluctuations due to transfer. Because HT implies opposite effects on the donor  and   recipient sub-populations, the integrals in $W^h$ appear in both equations with opposite signs.
This result has statistical implications. For instance, confidence intervals for $X^K_t$, $Y^K_t$ or confidence ellipsoids for $(X^K_t,Y^K_t)$ can be constructed from the variance of $\eta^K$, which can be computed from the above expressions. 

\subsection{Stochastic fluctuations for `quasi critical' populations on a long timescale}\label{sec:diffusion}

Diffusion approximations are obtained by accelerating time in `quasi critical' populations, in which the intrinsic growth rates (birth rate minus death rate) and the horizontal flux rate are of order $1/K$. As we let $K\rightarrow +\infty$, changes in the population are apparent only when considering the time scale $Kt$. This is similar to the renormalization leading to the classical Wright-Fisher diffusion.\\

Let $\gamma(.)$, $\nu(.)$, $\rho(.)$ be continuous positive functions. Let us consider a population where an individual with trait $u$ in the population $(x,y)\in (\N/K)^2$ has birth rate $\, \gamma(u)+\frac{\nu(u)}{K}$ and death rate
$$ \gamma(u)+\frac{\rho(u)}{K}+\frac{C(u,A)x}{K}+\frac{C(u,a)y}{K}.$$
The transfer rate is of the form
\begin{equation} K h_K(u,v,x,y)=\frac{\zeta+\frac{1}{K}\theta(u,v)}{\beta+\mu(x+y)}\label{coef_diffusion}
\end{equation}where $\zeta$ is a positive constant and $\theta$ can be positive or negative.

Under these assumptions, the sequence of stochastic processes $(X^K_{K.},Y^K_{K.})$ converges, in the time scale $Kt$ to the solution of the stochastic differential equations:
\begin{align}\label{eq:diffu}
\bar{X}_t= & x_0+ \int_0^t \Big[\big(\nu(A)-\rho(A)-C(A,A)\bar{X}_s-C(A,a)\bar{Y}_s\big)\bar{X}_s + \frac{\theta(A,a)-\theta(a,A)}{\beta+\mu(\bar{X}_s+\bar{Y}_s)}\bar{X}_s\bar{Y}_s\Big]ds \nonumber\\
&+ \int_0^t \sqrt{2 \gamma(A)\bar{X}_s} dB^A_s  +  \int_0^t \sqrt{\frac{2\zeta\bar{X}_s\bar{Y}_s}{\beta+\mu(\bar{X}_s+\bar{Y}_s)}}dB^h_s,\\
\bar{Y}_t= & y_0+ \int_0^t \Big[ \big(\nu(a)-\rho(a)-C(a,A)\bar{X}_s-C(a,a)\bar{Y}_s\big)\bar{Y}_s - \frac{\theta(A,a)-\theta(a,A)}{\beta+\mu(\bar{X}_s+\bar{Y}_s)}\bar{X}_s\bar{Y}_s\Big]ds\nonumber\\
&+ \int_0^t \sqrt{2\gamma(a)\bar{Y}_s} dB^a_s -  \int_0^t \sqrt{\frac{2\zeta\bar{X}_s\bar{Y}_s}{\beta+\mu(\bar{X}_s+\bar{Y}_s)}} dB^h_s,\nonumber
\end{align}where $B^A$, $B^a$ and $B^h$ are three independent Brownian motions which respectively capture the stochasticity of the birth and death processes in the $A$ sub-population, the stochasticity of the birth and death processes in the $a$ sub-population, and the stochasticity of the transfer process. In these equations, the integrals in $ds$ correspond to the regular (finite variation) part, due to the terms of order $1/K$ in the birth, death, and transfer rates. These terms become visible only because we consider the time scale $Kt$. The stochastic integrals with respect to $B^A$, $B^a$ and $B^h$ correspond to the irregular variations created by the succession of very rapid birth, death and transfer events, due to the constant part of the rates (i.e. the part that is independent of $1/K$). \\

\bi {\bf Link to the Wright-Fisher diffusion approximation of population genetics.} In order to  link these results with the Wright-Fisher model of population genetics, we first rewrite the diffusion approximation in \eqref{eq:diffu} in terms of the total population size $N_{s}=\bar{X}_s+ \bar{Y}_s$ and frequency of the traits
$P_{s} = \bar{X}_s/(\bar{X}_s+ \bar{Y}_s)$:
\begin{align}
N_t= & N_0+\int_0^t \Big\{(\nu(A)-\rho(A))P_s+(\nu(a)-\rho(a))(1-P_s)\nonumber\\
& -N_s\Big(C(A,A)P_s^2+C(a,a)(1-P_s)^2+(C(A,a)+C(a,A))P_s(1-P_s)\Big)\Big\} N_s\ ds \nonumber\\
& +\int_0^t \sqrt{2 N_s \big(\gamma(A) P_s + \gamma(a) (1-P_s) \big)} d\widetilde{B}_s 
\\
P_t= & P_0+\int_0^t \Big\{P_s(1-P_s)\Big[(\nu(A)-\rho(A))-(\nu(a)-\rho(a))\nonumber\\
& +N_s\Big((C(a,A)-C(A,A))P_s+(C(a,a)-C(A,a))(1-P_s)+\frac{(\theta(A,a)-\theta(a,A))}{\beta+\mu N_s}\Big)\nonumber\\
& -\frac{2}{N_s}\big(\gamma(A)-\gamma(a)\big)\Big] \Big\} ds\nonumber \\
& + \int_0^t \sqrt{\frac{2P_s(1-P_s)}{N_s} \left(\gamma(A) (1-P_s)  +\gamma(a)P_s+\frac{\zeta N_s}{\beta+\mu N_s} \right)}
d\widetilde{B}_s\label{diffusionP}
\end{align}
where $\widetilde{B}$ is a Brownian motion.\\

Assuming $\gamma(A)=\gamma (a)=\gamma $, the total demographic rate $2\gamma$ is the same for both traits. Then \eqref{diffusionP} writes
\begin{align}
\label{samegamma}
P_t= & P_0+\int_0^t \Big\{P_s(1-P_s)\Big[(\nu(A)-\rho(A))-(\nu(a)-\rho(a))+\frac{\theta(A,a)-\theta(a,A)}{\beta +\mu N_s}N_s\nonumber\\
& \hspace{2cm}+N_s\big((C(a,A)-C(A,A))P_s+(C(a,a)-C(A,a))(1-P_s)\big)\Big]\Big\}ds\nonumber\\
& + \int_0^t \sqrt{2\big( \gamma + \zeta\frac{N_s}{\beta+\mu N_s}\big)\,\frac{P_s(1-P_s)}{N_s}}d\widetilde B_s.
\end{align}We recover a generalization of the Wright-Fisher diffusion for $P_t$, with the classical term $P_s(1-P_s)/N_s$ in the variance.  In the case of frequency dependence with $\beta=0$ and $\mu=1$, the variance of the stochastic integral with respect to $\widetilde{B}$ reduces to $2 (\gamma+\zeta) P_s(1-P_s)/N_s$ where the effect of variation due to demography and transfer are additive and contribute equally. In the case of density-dependence with $\beta=1$ and $\mu=0$, the factor in front of $P_s(1-P_s)/N_s$ is $2\big( \gamma + \zeta\,N_s\big)$; as a consequence, depending on the population size $N_s$, the variance due to transfer can be negligible or very large compared to the variance due to the birth and death process. A similar result is true for BDA. When transfer rates are FD, HT makes the same quantitative contribution to genetic drift as demographic stochasticity.\\

If additionally, the competition kernel $\,C\,$ is assumed constant and the transfer almost critical and unilateral  ($\zeta=0$,   $\theta(a,A)=0$, and $\theta(A,a)=\theta \neq 0$), we obtain
\begin{align} \label{simplified_diffusion}
P_t= & P_0+\int_0^t \Big\{P_s(1-P_s)\Big[(\nu(A)-\rho(A))-(\nu(a)-\rho(a))+\frac{\theta}{\beta +\mu N_s}N_s \Big]\Big\}ds\nonumber\\
& + \int_0^t \sqrt{2 \gamma \,\frac{P_s(1-P_s)}{N_s}}d\widetilde B_s.
\end{align}
We recover the equation established by Tazzyman and Bonhoeffer \cite{tazzymanbonhoffer}, who specifically studied
the dynamics of plasmid transfer under the assumptions that time was discrete, population size was fixed, and HT was unilateral (only $A$ could be transferred to $a$ and not the reverse). Equation \eqref{samegamma} provides a generalization in which population size is dynamical, transfers are bilateral, and transfer rates can assume any of the general forms \eqref{eq:freqdeprate} to \eqref{eq:MMrate}.

\section{Probability and time of invasion and fixation under competition with horizontal transfer}\label{section:invasion}

In this section, we investigate the fate of a newly introduced individual with trait $A$ in a resident population in which trait $a$ is common; introduction of trait A may be due to mutation or migration. We assume that the invasion fitness of trait $A$ is positive, $S(A,a)>0$. According to Table \ref{tab:fix}, this includes both cases of an advantageous trait ($f(A,a)>0$), or a deleterious trait ($f(A,a)<0$) provided that the HT rate from $A$ to $a$ is high enough. Under those assumptions, the stochastic dynamics can be decomposed in up to three phrases, as illustrated in
 Fig.
 \ref{fig:fix}. The first phase begins with the introduction of an individual $A$ in the population, and ends when the size of the $A$ population either reaches a
 threshold $\varepsilon$ (i.e. the first time $t$ when $N^{A,K}_t>\varepsilon \, K$) or vanishes. If trait $A$ goes to fixation, the second phase can be approximated by the dynamical system given by \eqref{eq:edo} and has a duration of order $O(1)$. The third phase begins when the size of the $a$ population reaches a threshold $\varepsilon$, and ends when $a$ is lost.  If both traits $A$ and $a$ stably coexist, there is coexistence during a time of exponential order in $K$.   In all cases, the system goes to extinction on a time scale of exponential order in $K$.

 \begin{figure}[!ht]
 \begin{center}
\begin{tabular}{cc}
\includegraphics[height=6cm,trim =15mm 60mm 25mm 80mm, clip, scale=0.5]{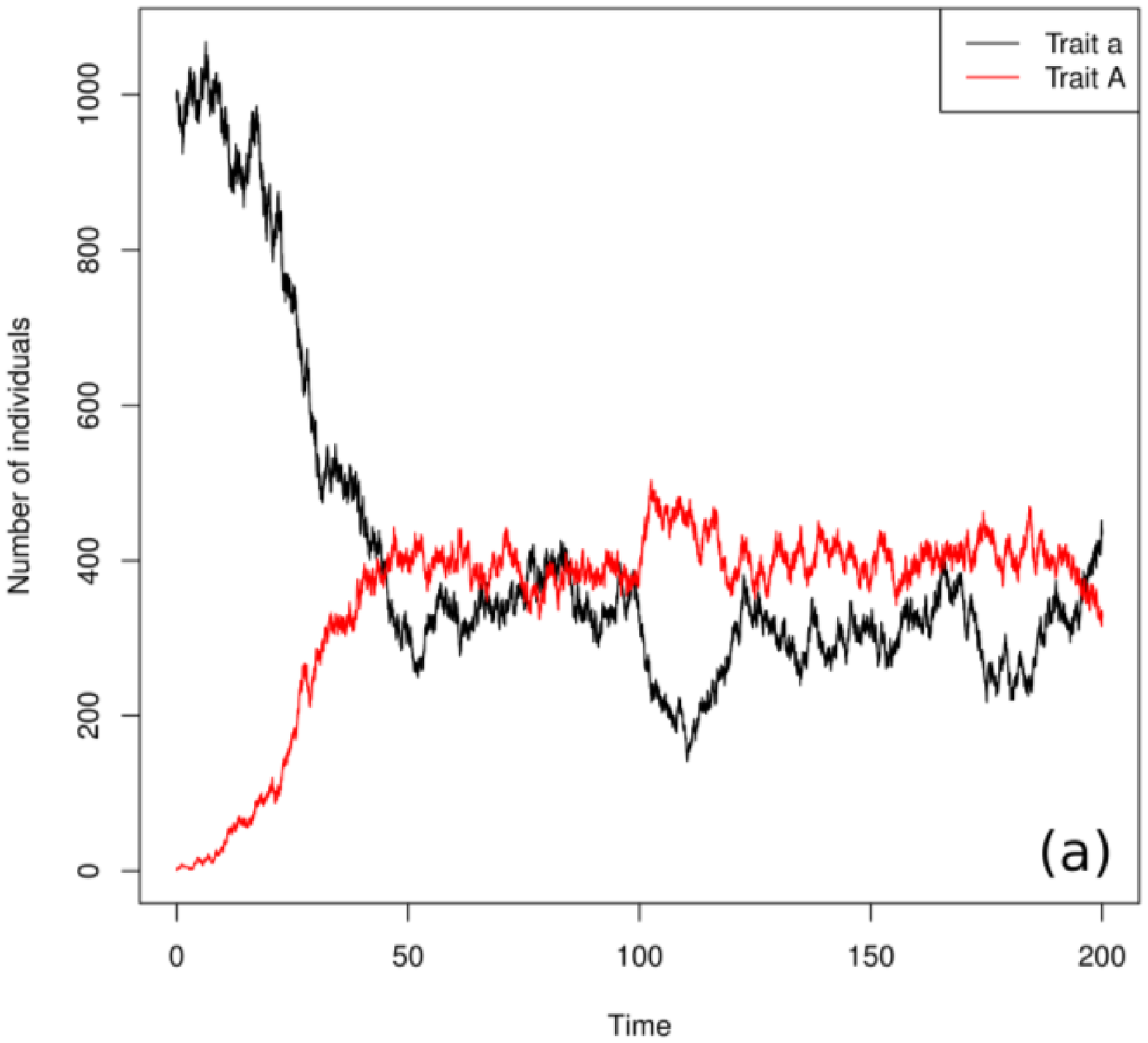} &
\includegraphics[height=6cm,trim =15mm 60mm 25mm 80mm, clip, scale=0.5]{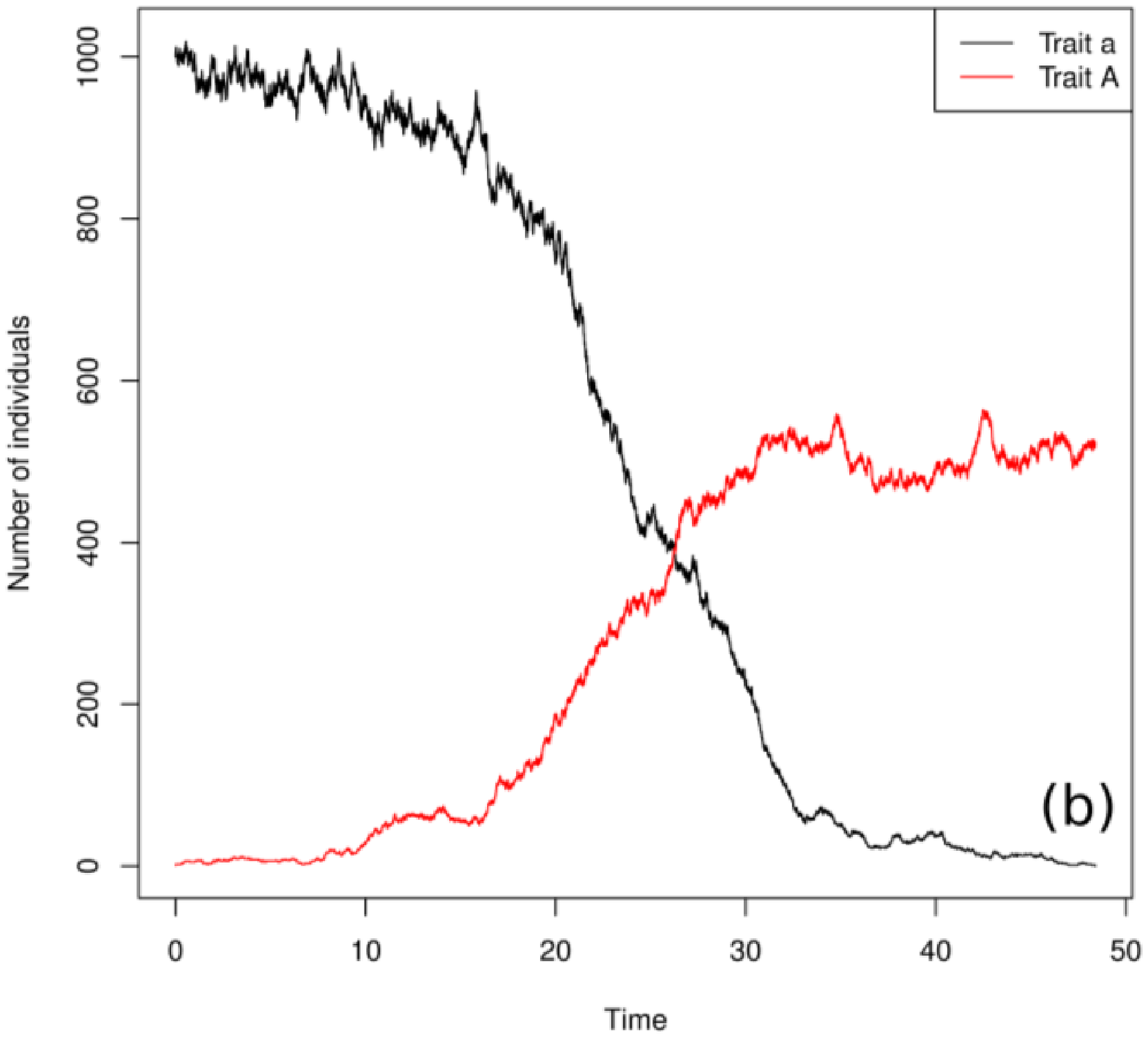} \\
\includegraphics[height=6cm,trim =15mm 60mm 25mm 80mm, clip, scale=0.5]{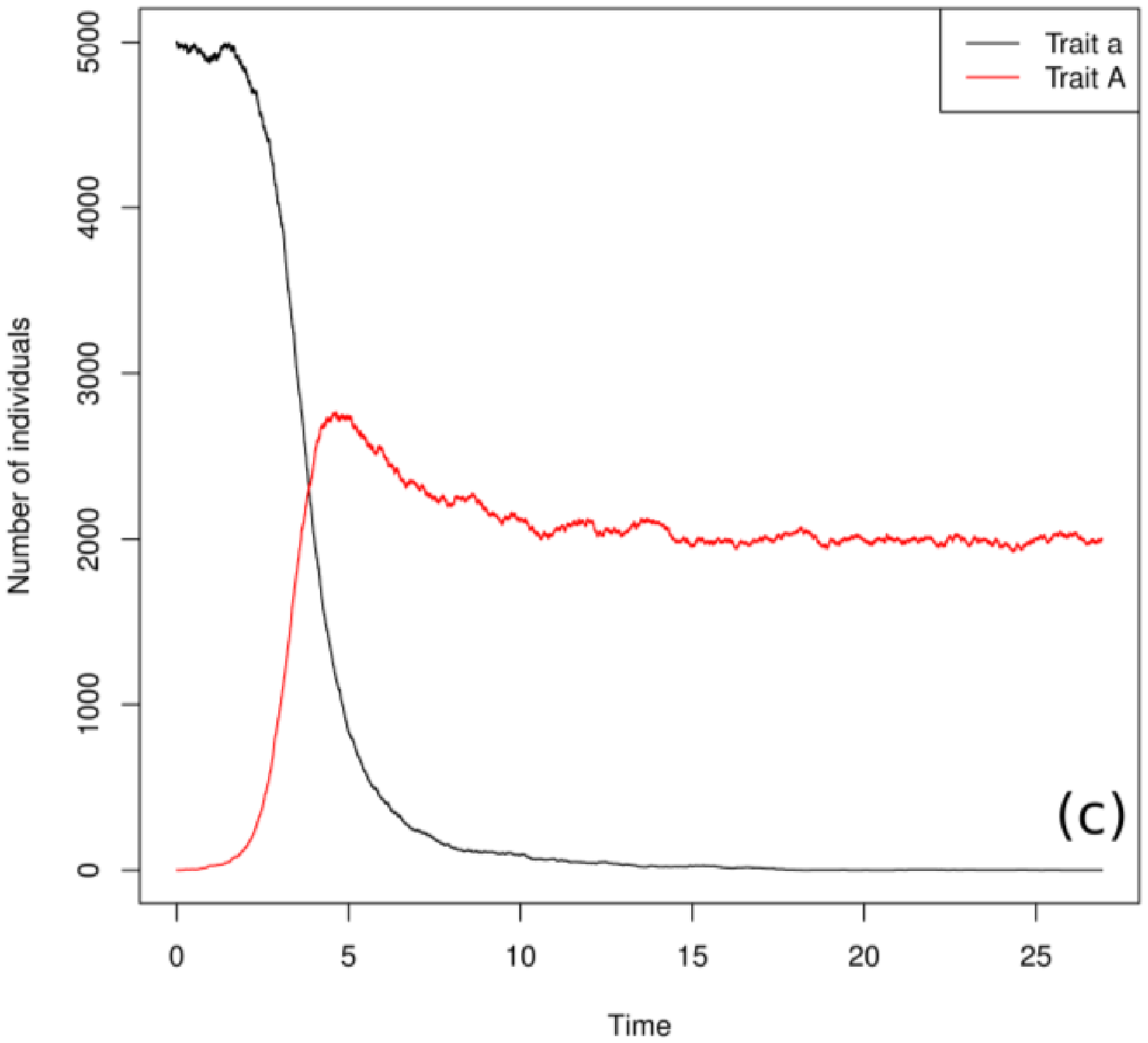} &
\includegraphics[height=6cm,trim =15mm 60mm 25mm 80mm, clip, scale=0.5]{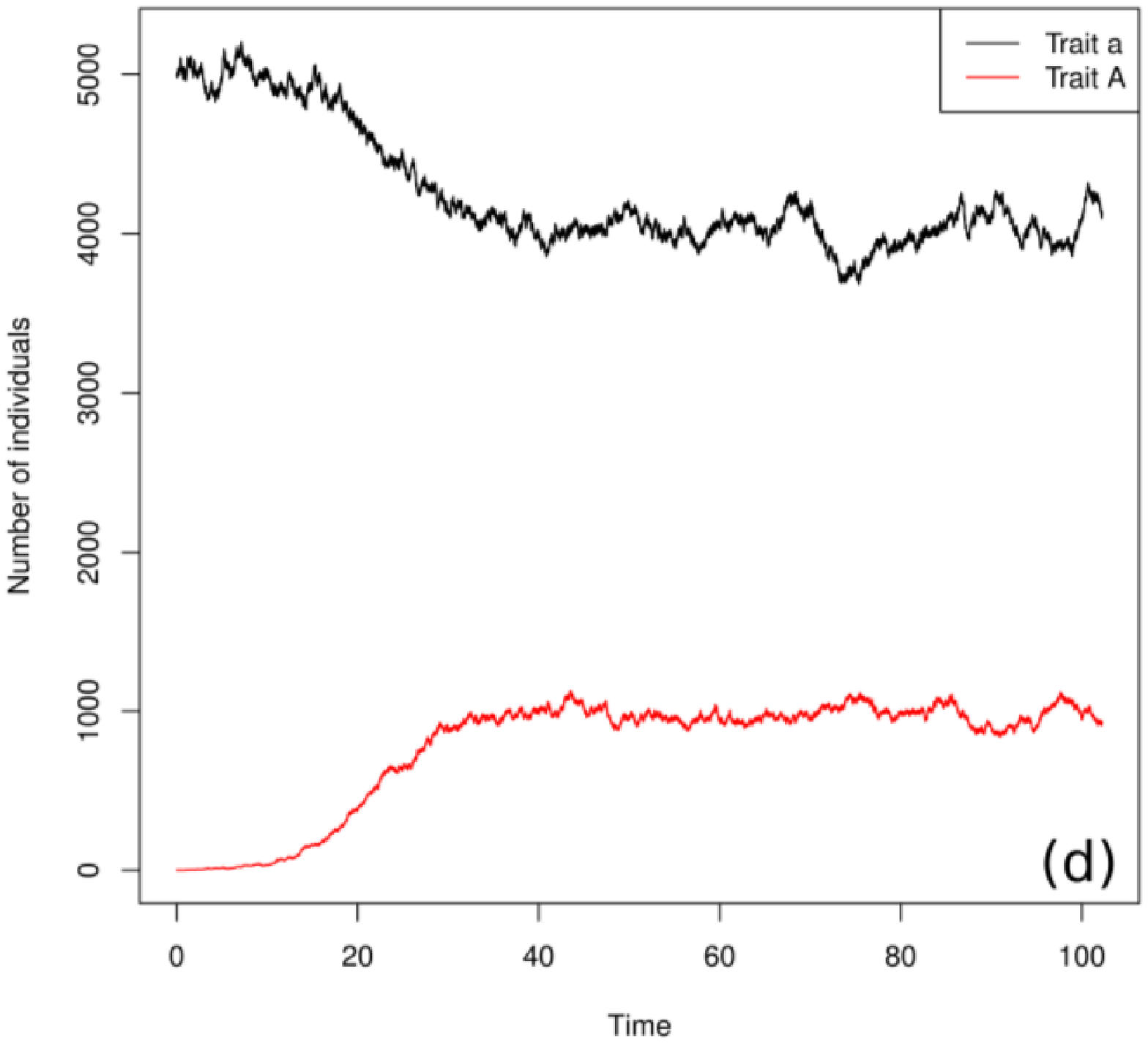} \\
\end{tabular}
\end{center}
\caption{{\small \textit{Invasion and fixation or polymorphic persistence of a deleterious mutation with density-dependent (left, (a) and (c), $\mu=0$, $\beta=1$) or frequency-dependent (right, (b) and (d), $\mu=1$, $\beta=0$,) unilateral HT rates. The deleterious nature of the mutation means that its invasion fitness without HT is negative. Other parameters: Top figures (a) and (b): constant competition coefficients $C(A,a)=C(a,A)=C(a,a)=C(A,A)=1$, $b(A)=0.5$, $b(a)=1$, $d(a)=d(A)=0$ $K=1000$, $\alpha =0.7$; Bottom figures (c) and (d): $C(A,a)=C(a,a)=2$, $C(A,A)=4$, $C(a,A)=1$, $b(A)=0.8$, $b(a)=1$,  $d(a)=d(A)=0$ $K=10000$,  $\alpha =5$ under density-dependent rate, $\alpha=0.5$ under frequency-dependent rate.}}}
\end{figure}\label{fig:fix}

\subsection{Probability of invasion and fixation}

During the first phase, $N^{A,K}$ can be approximated by a linear birth-death branching stochastic process, which shows that the phase ends with $X^K$ reaching the threshold $\varepsilon$ with probability (e.g. \cite{champagnat06,champagnatferrieremeleard})
\begin{equation}
P(A,a)=\frac{S(A,a)}{b(A)+ h(A,a,0,\overline{y})\,\overline{y}}= \frac{b(A)- d(A) + \big( \frac{\alpha(A,a)}{\beta+\mu \overline{y}} -  C(A,a)\big)
\,\overline{y}}{b(A)+ \frac{\tau(A,a) \,\overline{y}}{\beta+\mu \overline{y}}}.\label{eq:proba_invasion}
\end{equation}
In Table \ref{tab:fix}, this probability of invasion is expounded for each form of HT rate.
Recall that without transfer, the probability of invasion is $\pi(A,a):= [f(A,a)]_+/b(A)$ where $f(A,a)=r(A)-C(A,a)\,
\overline{y}$ is the fitness function.\\

Comparing the probability of invasion with and without transfer, \eqref{eq:proba_invasion} shows that HT increases the probability of
invasion of a mutant if
\ben \label{eq:criteria}
\frac{f(A,a)}{b(A)}<1-\frac{ \tau(a,A)}{ \tau(A,a)}.
\een
If transfer is symmetrical ($\tau(a,A) = \tau(A,a)$) this condition is always satisfied for a deleterious mutation and never satisfied for a beneficial mutation. Thus, symmetrical transfer always facilitates the invasion of a deleterious mutation and always hampers invasion of a beneficial mutation. The latter is because HT  increases stochasticity and variance of population fluctuations: a beneficial trait $A$ that just appeared in a population can not only be lost because of the death of the newly mutant, but also because of a HT event from an $a$ individual to the initial $A$ individual.  \\

From Section \ref{section:ODE} we know that invasion does not necessarily imply fixation, even when the invasion fitnesses of the two types have opposite sign, as shown by Fig. \ref{fig:diagram} (5) and (6). In these cases, fixation depends on initial conditions and is usually not achieved when the invading type starts from a small density. Considering the special case of constant competition ($C(u,v)\equiv C$), however, invasion does imply fixation (Fig. \ref{fig:diagram} (1)) if HT rates are FD or when condition \eqref{eq:DDpolym} is not satisfied if HT rates are DD or BDA. In these cases, the probability of fixation is equal to the probability of invasion, given by Table \ref{tab:fix}.

\begin{table}[!ht]
    \centering
\small
\begin{tabular}{|l|c|c|}
\hline
Transfer rate model  & Invasion fitness $S(A,a)$ & Invasion probability $P(A,a)$ \\
\hline
DD : $\tau(A,a)$ & $f(A,a)+ \frac{\alpha(A,a) r(a)}{C(a,a)}$ &  $\frac{\big[f(A,a)+\frac{\alpha(A,a)r(a)}{C(a,a)}\big]_+}{b(A)+\frac{\tau(A,a)r(a)}{C(a,a)}}$  \\
FD : $\frac{\tau(A,a)}{x+y} $ & $f(A,a)+\alpha(A,a)$ &$\frac{\big[f(A,a)+\alpha(A,a)\big]_+}{b(A)+\tau(A,a)}$   \\
BDA : $\frac{\tau(A,a)}{\beta+\mu(x+y)}$ &$f(A,a)+\frac{\alpha(A,a) r(a)}{\beta C(a,a) + \mu r(a)}$ &$\frac{\big[f(A,a)+\frac{\alpha(A,a)r(a)}{\beta C(a,a)+\mu r(a)}\big]_+}{b(A)+\frac{\tau(A,a)r(a)}{\beta C(a,a)+\mu r(a)}}$ \\
\hline
\end{tabular}
 \caption{{\small \textit{Invasion fitness and invasion probabilities for each model of transfer rates. DD and FD are special cases of BDA with $\beta=1$, $\mu=0$ and $\beta=0$, $\mu=1$ respectively. Without transfer, the invasion fitness and invasion probability are  $f(A,a)=r(A)-\frac{r(a)C(A,a)}{C(a,a)}$ and $\pi(A,a)=[f(A,a)]_+/b(A)$}}}\label{tab:fix}
\end{table}

\subsection{Times of invasion and fixation}

As the selectively advantageous trait A increases from rare, the first phase of $A$ population growth has a duration of order $\log K/S(A,a)$. If $X^K$ reaches the threshold $\varepsilon$, then the second phase begins, where the processes $(X^K,Y^K)$ stay close to the dynamical system \eqref{eq:edo}.  The deterministic trajectory, which has a duration of order 1,  can reach one of two final states: either both types of individuals stably coexist, or individuals with trait $A$ invade the population and the $a$ population reaches the threshold $\varepsilon$ (i.e. $N^{a,K}_t< \varepsilon \,K$). Should the latter happens, the third phase begins and $N^{a,K}$ can be approximated by a linear birth-death branching process, until $A$ is fixed and $a$ is lost.
In this birth-death process, the transfer $A \to a$ acts as a birth term and the transfer $a\to A$ as a death term.
The third phase has an expected duration $\E_{\varepsilon \,K} \left[ T_0 \right]$ of  (see \cite[Section 5.5.3, p.190]{Meleardbook})
\ben
\E_{\varepsilon \,K} \left[ T_0 \right]=\frac{1}{b} \sum_{j\geq 1} \Big( \frac{b}{d}\Big)^j \ \sum_{k=1}^{\varepsilon K -1}\frac{1}{k+j}
\een
where
\begin{align*}
b= & \ b(a)+\frac{\tau(a,A)r(A)}{\beta C(A,A)+\mu r(A)}\\
d= & \ d(a)+\frac{C(a,A)r(A)}{C(A,A)}+\frac{\tau(A,a)r(A)}{\beta C(A,A)+\mu r(A)}.
\end{align*}Note, this is a case where the intensities of directional transfers, as measured by $\tau(A,a)$ and $\tau(a,A)$, matter - not just the flux $\alpha(A,a)$.
When $K \rightarrow \infty$, $\E_{\varepsilon \,K} \left[ T_0 \right] \simeq \frac{\log K}{d-b}$, which means that the third phase is of order $\log K/|S(a,A)|$ in duration. Summing up, the fixation time of an initially rare trait $A$ going to fixation is of order
\be \label{eq:Tfix}
T_{fix}=\log K /S(A,a)+1+\log K /|S(a,A)|, 
\ee
where the expressions for $S(A,a)$ and $S(a,A)$ are given in Table \ref{tab:fix}.\\

Equation \eqref{eq:Tfix} shows that if the HT rate is biased towards the transfer of $A$ to $a$ ($\alpha(A,a)>0$), then the fixation time decreases with $\alpha(A,a)$. In the DD and BDA cases, this effect is amplified by a larger value of the equilibrium population sizes $\bar{y}=r(a)/C(a,a)$ and $\bar{x}=r(A)/C(A,a)$.

\subsection{Case of unilateral horizontal transfer, such as for plasmids}\label{sec:plasmid}

In this section, we focus on the special case of unilateral transfer, which has been used to address the question of fixation of mobile genetic elements such as plasmids, for instance. Plasmid transfer is unilateral: individuals containing a specific plasmid can transmit one copy to another individual which does not carry this plasmid. Let us assume that trait $A$ indicates that the individual carries the plasmid of interest; individuals with trait $a$ are devoid of this plasmid. Unilateral transfer then means $\tau(A,a)>0$ and $\tau(a,A)=0$, hence $\alpha(A,a)=\tau(A,a)$.

Unilateral transfer has been modelled in a stochastic two-type population genetics framework by Novozhilov et al. \cite{novozhilovetal2005} and Tazzyman and Bonhoeffer \cite{tazzymanbonhoffer}. These studies focused on FD transfer rates, and assumed constant population size (Novozhilov et al. used a Moran's model, and Tazzyman and Bonhoeffer used a Wright-Fisher model with no overlapping generation). To compare our results with theirs, we focus on constant competition coefficients ($C(u,v)\equiv C$) in the rest of the section.

\paragraph{Invasion and fixation of a plasmid}

By definition,  invasion of $A$ into $a$ requires  $S(A,a)>0$. According to Table \ref{tab:fix}, invasion occurs under BDA if
\begin{equation} \tau(A,a)> -\Big(\frac{\beta}{ \bar{y}}+\mu \Big)f(A,a),\label{unilateral:invasionDD}
\end{equation}
where $\bar{y}$ is the equilibrium size of a monomorphic population $a$.
Invasion is always possible if $f(A,a)>0$ but for $f(A,a)<0$, the plasmid can only invade if the transfer rate is high enough. Under FD ($\beta=0$), the invasion of the plasmid does not depend on the equilibrium population size. Under DD and BDA ($\beta >0$), the larger the resident population $a$, the easier the invasion of the plasmid.  \\

We saw that FD transfer rates and constant competition result in invasion implying fixation. Under these conditions, Novozhilov et al. \cite{novozhilovetal2005} found that the probability of fixation is $f(A,a)+\tau(A,a)$; Tazzyman and Bonhoeffer \cite{tazzymanbonhoffer} found an additional two-fold factor due to the difference between Moran and Wright-Fisher models. Tazzyman and Bonhoeffer concluded that horizontal transfer and vertical transmission of a trait under selection have similar effects on the fate of the trait, hence on  the adaptation process. Our results show that this it is not true if the influence of ecological competition on population size is taken into account. Our model predicts the probability of fixation to be
\begin{equation}
\frac{f(A,a)+\tau(A,a)}{b(A)+\tau(A,a)}.\label{profixCct}
\end{equation}
The probability of fixation of trait $A$ thus increases linearly with its fitness $f(A,a)$ through vertical transmission, in accordance with previous results (\cite{novozhilovetal2005,tazzymanbonhoffer}). In contrast, the probability of fixation grows in a decelerating and saturating manner with the HT rate $\tau(A,a)$; the relationship becomes closer to the linear one predicted by \cite{novozhilovetal2005,tazzymanbonhoffer} only when the horizontal transmission rate is small relatively to the birth rate. The slope of \ref{profixCct} with respect to $\tau(A,a)$ near zero is equal to $$\frac{b(A)-f(A,a)}{b(A)^2}$$and thus can be large in organisms with low birth rate. For such organisms (and under the current assumptions of unilateral FD transfer and constant competition), the probability of fixation of an advantageous trait $A$ will be more sensitive to the HT rate $\tau(A,a)$ than to the selective advantage. To conclude,  it is possible for HT to have major effects on the distribution of mutational effects that are fixed and contribute to adaptation (see e.g. \cite{orr1998, eyrewalkerkeightley2007}).

\paragraph{Case of costly plasmids}

Mobile genetic elements such as plasmids generally impose a high cost to the carrier \cite{baltrus2013},
 \emph{i.e.} $r(A)<r(a)$. Without HT, the invasion fitness would be $f(A,a)=r(A)-r(a)<0$, and the deleterious trait $A$ cannot be maintained. We investigate the conditions under which HT can facilitate the invasion, maintenance and fixation of a costly plasmid. The possible dynamics are illustrated in Fig. \ref{fig:diagram}.\\

Equations \eqref{unilateral:invasionDD} show that however large the reproductive cost ($f(A,a)<0$) 
 of the plasmid, invasion will always happen provided the transfer rate is high enough. How high the transfer rate can be in reality may be constrained by additional factors such as the intrinsic cost of transfer events.\\


The invading plasmid will then go to fixation under FD (cf Section \ref{section:ODE}). Under DD,  the possibility of polymorphism maintenance of the plasmid  was shown  in \cite{stewartlevin1977}. Our model provides an explicit condition (see \eqref{eq:DDpolym}): $$-\frac{f(A,a) C}{r(a)}< \tau(A,a) < -\frac{f(A,a) C}{r(A)}\Longleftrightarrow-\frac{f(A,a)}{\bar y}< \tau(A,a) < -\frac{f(A,a) }{\bar x}.$$ When the transfer rate is too small, the costly plasmid may not invade, as noted above.  When the transfer rate is too high,  the plasmid invades and goes to fixation. Similar results can be obtained in the general BDA case.\\

A necessary (but not sufficient) condition for the fixation of a plasmid is $S(a,A)<0$ and $S(A,a)>0$; in Fig. \ref{fig:diagram} this corresponds to cases (1) and (6). The realization of these conditions is favored by a large transfer rate $\tau(A,a)$. However, the final outcome - fixation or polymorphism - may depend on the initial density of plasmids. This can be seen in case (6) of Fig. \ref{fig:diagram} where fixation only occurs if the initial density of plasmid carriers is high enough. This shows that costly traits can be maintained in polymorphism even in absence of spatial structure or frequency-dependent selection.

\section{Overview and concluding remarks}

We have constructed a model for the dynamics of two interacting populations, each being characterized by a `trait' which is inherited vertically (under the assumption of clonal reproduction) and can be exchanged horizontally upon contact between individuals. The `traits' can describe genes, plasmids, endosymbionts, or cultural information; they may influence the birth and/or death rates of their bearers, as well as the intensity of ecological competition among them. We called $A$ and $a$ the two values or states of the traits. Starting from a `microscopic' description of stochastic birth, death, and contact events  at the level of individuals, we first derived a general model for the rate of contact (BDA), of which frequency-dependent (FD) and density-dependent (DD) rates are special cases (cf. Section \ref{section:modele}). This extends previous studies and discussions of contact processes in epidemiology (see \cite{begon,diekmannheesterbeek,mccallum,turner}) and provides a unifying mathematical validation for the notions of frequency-dependent versus density-dependent contact rates - both models can be recovered by taking different limits on the same underlying stochastic individual-level process. Whereas McCaig et al. \cite{mccaig} took a cybernetic (algorithmic) approach to the same problem (i.e. scaling up from individual-level interactions to population-level transmission models), our approach provides an analytical treatment in which stochastic processes are modeled explicitely. The mathematical limits by which transfer rates are derived lead us to expect density-dependent HT rates when the population size is low, and frequency-dependent HT rates when the population is close to its carrying capacity. Although measuring the transfer rates of genes, plasmids or endosymbionts remains a major empirical challenge \cite{philipsen,soren,zhong}, there are some observations for plasmids that do suggest such a correlation between the form (density- versus frequency-dependent) of the transfer rate and the state of the population (low size versus close to carrying capacity) (Raul Fernandez-Lopez, pers. com.).\\

Taking a large-population limit on the stochastic individual-level model, we obtained a deterministic model which takes the form of a Lotka-Volterra competition system with additional terms accounting for HT (Section \ref{section:ODE}, equations \eqref{eq:edo}). The stability analysis of this system revealed the possible patterns of invasion of one trait by the other, or coexistence of both traits. From this analysis, we calculated invasion fitness taking HT into account; three conclusions followed: \\
(i) HT can revert the direction of selection, i.e. invasion fitness ($S$) and selective value ($f$) can have opposite signs. A necessary condition is that the transfer flux (negative or positive) more than compensate for the selective value (advantage or disadvantage) of the rare trait, and a smaller resident population makes the condition more likely to be sufficient. Thus, HT can drive invasion of a deleterious trait, or prevent invasion of an advantageous trait.\\
(ii) Invasion does not necessarily imply fixation, even if the traits' invasion fitnesses have opposite signs, and even if their phenotypic effects are small. Thus, HT causes violation of the otherwise general `attractor inheritance' principle of Geritz et al. \cite{geritzgyllenbergetal}. Due to HT, both traits may coexist in a stable polymorphism even if their invasion fitnesses are of opposite sign.\\
(iii) Polymorphic coexistence may occur even when both invasion fitnesses are negative, i.e. neither trait is able to grow from rarity in a resident population of the other trait. This requires that the initial population contains both traits at sufficiently high frequencies.\\

In the case where the traits have no effect on the competition coefficients, the classical Lotka-Volterra model predicts exclusion of one type by the other, whereas with HT, automatic exclusion of one type by the other is the rule only in the special case of FD transfer rates. Our deterministic model may also be compared to epidemiological models of disease transmission in which two classes (susceptibles and infectives) are distinguished and host demographics account for resource competition (a seminal contribution in this vein was Gao and Hethcote \cite{gao}, see also \cite{lili} in the case of plasmid transmission). Epidemiological theory has highlighted the importance of the effect of host state (susceptible versus infected) on host intraspecific competition; ``emergent carrying capacity'' models thus recognize infection-modified host competitive abilities (see \cite{sieber}). Here the case of frequency-dependent transfer rates highlights that trait-dependent competitive abilities can lead to very different dynamical behaviors, including the possibility of polymorphism (i.e. stable coexistence of both traits) and even bistability between an exclusion equilibrium (only one type present at equilibrium) and a polymorphic equilibrium (see Fig. \ref{fig:diagram} (5)-(6)). Bistability of polymorphic equilibria and tri-stability among both exclusion equilibria and one polymorphic equilibrium become possible under the more general form of transfer rates (BDA). The likelihood of these dynamical scenarios (Cases (5)-(8) in Fig. \ref{fig:diagram}) is very small when parameters are drawn at random; however, the fact that they are possible for traits with small effects on phenotypes (see Special case 2 in Section \ref{section:ODE}) calls for studying their attainability by adaptive evolution proceeding as a trait substitution sequence (\cite{metzgeritzmeszenajacobsheerwaarden} and \cite{champagnatferrieremeleard}). In other words, evaluating the biological significance of these dynamical scenarios requires that we determine their evolutionary attractivity and stability - an open question that we are currently investigating.\\

What is the effect of demographic stochasticity and stochasticity of transfer events on the population dynamics predicted by the deterministic model (Section \ref{sec:fluctuation})? We found that the effect of HT stochasticity on population fluctuations is not determined solely by the net transfer flux ($\alpha(A,a)$), but is influenced by the sum of transfer rates in both directions ($\tau(A,a)$ and $\tau(a,A)$); thus fast transfers that balance out (small $\alpha$) may nonetheless cause large stochastic fluctuations in the size of both subpopulations. In the case of `quasi-critical' populations that have very small growth rate and transfer flux, we focused on the case of demographically neutral traits and found that the contact process had a key influence on the relative effect of HT on population variance. With FD transfer rates, the effect of birth-death stochasticity and transfer stochasticity are additive and contribute equally to population variance, in line with the results of Tazzyman and Bonhoeffer \cite{tazzymanbonhoffer}. With DD transfer rates, the relative effect of transfer stochasticity can become very large or very small depending on the population size.\\

Assuming that one trait (e.g. $A$) is initially rare in a population of the other trait ($a$), we focused on the case of $A$ potentially invading (i.e. $S(A,a)>0$) and derived exact analytical expressions for the probability of invasion and time to fixation for each model of transfer rates (DD, FD, and the general case BDA) (Section \ref{section:invasion}). We derived the general condition for HT to increase the probability of invasion. In the case of symmetrical transfer, HT always increases the invasion probability of a deleterious trait and always decreases the invasion probability of a beneficial trait (due to the stochasticity of the transfer process). If $A$ goes to fixation, a bias of transfer in favor of $A$ will speed up fixation; under DD or BDA (not FD) transfer rates, the larger the resident population, the stronger this effect.\\

Finally we addressed the case of unilateral transfer, as for plasmids, assuming no effect of traits on competition coefficients (constant $C$). Unilateral HT does not alter the invasion potential of a beneficial trait; for deleterious traits, unilateral HT promotes invasion if the transfer rate is high enough, and invasion is facilitated in a larger resident population (provided that the transfer rate is DD or BDA, not FD). With FD transfer rates (and trait-independent competition), invasion implies fixation, and we found the probability of fixation to be not simply $f(A,a) + \tau(A,a)$ (as was found by \cite{tazzymanbonhoffer}) but $(f(A,a) + \tau(A,a))/(b(A) + \tau(A,a))$. Thus, vertical transmission and horizontal transfer are not equivalent in determining fixation; the probability of fixation of a beneficial trait ($f(A,a) > 0$) becomes more sensitive to the transfer rate $\tau(A,a)$ than to the selective value $f(A,a)$ in organisms in which vertical transmission is slower (i.e. smaller birth rate). In the case of costly plasmids (i.e. trait $A$ is deleterious), invasion is always possible provided that the transfer rate is large enough. Invasion implies fixation under FD, but under DD maintenance of the plasmid in a polymorphic population is possible, for intermediate values of the transfer rate.\\

In conclusion, HT interacts with ecology (competition) in non-trivial ways. Competition influences individual demographics, and this in turns affects population size (that we do not assume constant), which feeds back on the dynamics of transfers. This feedback loop has complex, previously unknown, effects on the dynamics of deleterious traits (including the case of costly plasmids), making their stable polymorphic maintenance possible, even in the absence of frequency-dependent selection, spatio-temporal heterogeneity, compensatory mutations \cite{dionisio}, mutation-selection balance \cite{novozhilovetal2005}, or imperfect horizontal transmission - selection balance \cite{jaenike2012} - all mechanisms which are classically invoked to explain stable polymorphisms \cite{slateretal2008}. The population-size mediated interaction between competition and transfer has other notable consequences, including (i) a strong contribution of transfer stochasticity (relative to demographic stochasticity) on population fluctuations when transfer rates are density-dependent or of the more general form BDA (ii) a greater acceleration of fixation by HT obtained by weakening competition in the resident population. Our modeling framework provides a basis to develop a general theory for the influence of HT on evolutionary adaptation, where trait variation may represent different types of transferrable elements (as in Mc Ginty et al. \cite{mcgintyetal2013} who studied the evolution of plasmid-carried public goods, and Doebeli and Ispolatov \cite{ispolatovdoebeli} who modelled the adaptive evolution and diversification of cultural ideas), or host phenotypes differing in their control of or response to transfer (as in Gandon and Vale \cite{gandonvale} who studied the evolution of resistance to foreign genetic elements), or both.

\section*{Supplementary materials}

See \cite{billiardcolletferrieremeleardtran:ESM}.

\section*{Author's contribution}
All authors designed the study and the models. P.C., S.M. and V.C.T made the mathematical proofs. All authors wrote the paper. All authors gave final approval for publication.
\section*{Competing interests}
We declare we have no competing interests.
\section*{Funding}
S.B., S.M. and V.C.T. have been supported by the ANR MANEGE (ANR-09-BLAN-0215), the Chair ``Mod\'elisation Math\'ematique et
Biodiversit\'e" of Veolia Environnement-Ecole Polytechnique-Museum National d'Histoire Naturelle-Fondation X. V.C.T. also acknowledges support from Labex
CEMPI (ANR-11-LABX-0007-01) and has been invited by the University of Arizona. R.F. acknowledges support from the P\'epini\`ere Interdisciplinaire CNRS-PSL ``Eco-Evo-Devo" and the Partner University Fund.

{\footnotesize
\providecommand{\noopsort}[1]{}\providecommand{\noopsort}[1]{}\providecommand{\noopsort}[1]{}\providecommand{\noopsort}[1]{}

}

\end{document}